\documentclass[letterpaper,10pt]{article} 

\usepackage{opticameet3} 

\newcommand\authormark[1]{\textsuperscript{#1}}

\usepackage{amsmath,amssymb}
\usepackage[colorlinks=true,bookmarks=false,citecolor=blue,urlcolor=blue]{hyperref} 
\usepackage{adjustbox}
\usepackage{multirow}
\usepackage{multicol}
\usepackage{diagbox}
\usepackage{textcomp} 

\begin{document}

\title{Experimental Demonstration of MPI-Penalty-Free S-band Transmission over G.654.E Fibres}

\author{Romulo~Aparecido\textsuperscript{(1)}, 
    Jiaqian~Yang\textsuperscript{(1)},
    John~D.~Downie\textsuperscript{(2)},
    Lidia~Galdino\textsuperscript{(2)},
    Eric~Sillekens\textsuperscript{(1)},
    Henrique Buglia\textsuperscript{(1)},
    Ronit Sohanpal\textsuperscript{(1)},
    Robert I. Killey\textsuperscript{(1)},
    and Polina Bayvel\textsuperscript{(1)}}

\address{\authormark{1} Optical Networks Group, University College London (UCL), London, UK\\
\authormark{2}Corning Research and Development Corporation, Corning, NY 14831, USA
}

\email{\authormark{*}romulo.aparecido.22@ucl.ac.uk} 

\begin{abstract}
We demonstrate multipath interference-penalty-free S-band transmission over two G.654.E-compliant fibres with cable cutoff up to 1530 nm. No SNR degradation was observed across various transmission distances and baud rates.
\end{abstract}


\vspace{-0.0cm}

\section{Introduction}

\vspace{-0.15cm}

Optical communication systems form the foundation of global data transmission~\cite{Winzer:18,bayvel2016maximizing}.
Given the largely unused ultra-wideband (UWB) spectrum available with optical fibres~\cite{9893169}, maximising the usage of wavelength bands is an attractive and potentially cost-effective approach to increase system throughput.
C+L band systems are currently being implemented in deployed networks and the inclusion of the S-band (1460~nm - 1530~nm) is foreseen to be one of the most likely candidates for future upgrades. 
Groundbreaking results have been reported using communication systems transmitting channels over SCL bands~\cite{Puttnam:22}, with the use of amplifying schemes tailored to the S-band, such as thulium-doped fibre amplifiers (TDFAs) and Raman amplifiers~\cite{9489895, 9144561}.

In parallel, new optical fibres have been developed to further increase the capacity and reach of optical systems.
Specifically, G.654.E-compliant fibres feature ultra-low-loss properties and a larger effective area (125~\textmu m$^2$) compared to the standard G.652.D fibres (80~\textmu m$^2$)~\cite{Maeda:21, Mlejnek:19}. 
This design reduces the impact of amplified spontaneous emission (ASE) noise due to the lower attenuation values, reaching 0.148 dB/km in 1550~nm. 
Additionally, the larger effective area reduces nonlinear effects such as inter-channel stimulated Raman scattering (ISRS), a phenomenon that can severely limit ultra-wideband (UWB) transmission by transferring power from shorter to longer wavelength channels~\cite{9748243, 8345898,101109ecoc2021}.
The increased effective area, however, results in a shift in the cable cutoff (CC) towards longer wavelengths of up to 1530~nm~\cite{itut}.
As a result, the coupling to higher-order modes may generate multi-path interference (MPI) and, in turn, reduce the throughput in the S-band.  
In a previous study, S-band transmission over G.654.E Corning\textsuperscript{\textregistered} TXF\textsuperscript{\textregistered} optical fibre was investigated by modelling the coupling between modes and the differential mode attenuation~\cite{downie2023ecoc,Lin:21}.
It was shown that even in the most adverse cases, where bend conditions representing loose-tube cables were considered, the MPI levels in the S-band are below -70~dB/km, which indicates a signal-to-noise ratio (SNR) penalty of less than 0.1~dB.
However, this has not been confirmed experimentally to show S-band penalty-free transmission over G.654.E-compliant fibre. 
In this work, we present the first experimental investigation of S-band transmission over Corning\textsuperscript{\textregistered} TXF\textsuperscript{\textregistered} and Corning\textsuperscript{\textregistered} Vascade\textsuperscript{\textregistered} EX2500 fibres, both of which are G.654.E-compliant.
The results demonstrate the absence of SNR degradation, showing that MPI is negligible and confirming the findings of the theoretical model~\cite{downie2023ecoc}. 
In addition, we observed significant SNR improvement of approximately 3~dB when transmitting S-band signals over 160~km of Vascade EX2500 compared to standard single-mode fibre (G.652.D).
This study shows the practicality of employing G.654.E fibres for S-band transmission. 
Not only do these not exhibit SNR penalties due to MPI, they can, in fact, enhance transmission performance due to their lower attenuation.

\vspace{-0.2cm}

\section{Experimental Setup}

\begin{figure*}[t]
    \centering
    \includegraphics[width=0.85\textwidth]{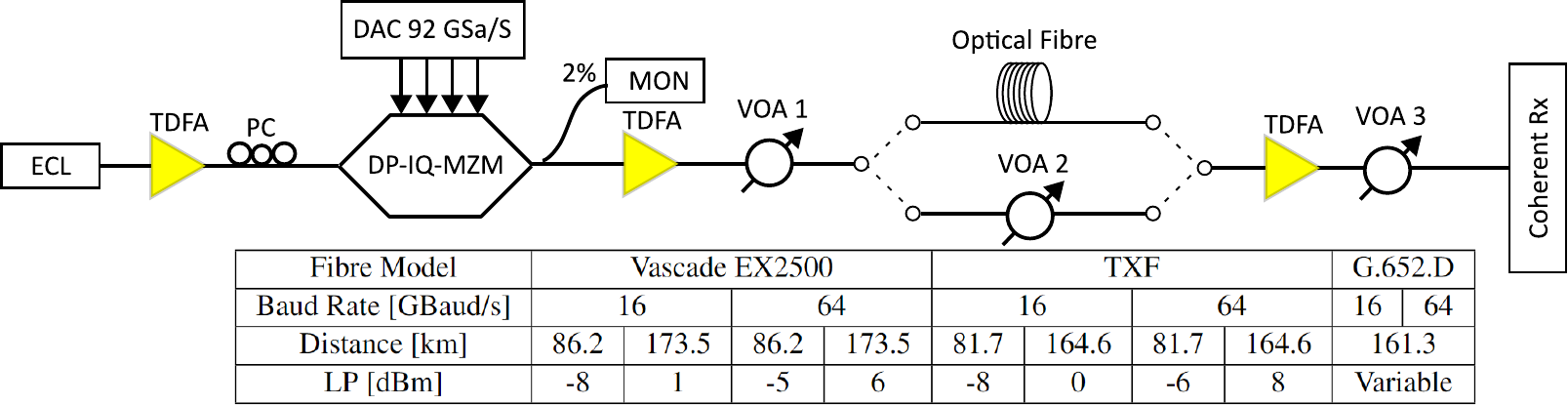}
    \setlength{\belowcaptionskip}{-0pt}
    \caption{Schematic diagram of the experimental setup to evaluate transmission performance in the S-band. The transmission over fibre is compared to an attenuation stage. The inset shows a table with baud rate, distance, and LP used for transmission over the different fibre models.}
    \label{fig:setup}
      
\end{figure*}

\vspace{-0.15cm}

Figure~\ref{fig:setup} depicts the experimental transmission setup used to assess SNR degradation from MPI in the S-band.
An external cavity laser (ECL) with \textless100~kHz linewidth was used to generate the carrier, further amplified by a booster TDFA.
A polarisation controller (PC) was used to adjust the signal polarisation state into the modulator.
The carrier was modulated by a 35~GHz 3-dB bandwidth electro-optical dual-polarisation in-phase quadrature Mach-Zehnder modulator (DP-IQ-MZM), driven by 92~GSa/s 8-bit digital-to-analog converters (DACs) to produce 16~QAM signals at baud rates of either 16~GBaud or 64~GBaud.
The optical launch power (LP) was adjusted by a variable optical attenuator (VOA~1), followed by an optical switch to route the light to either the optical fibre or a second VOA (VOA~2).
The optical switch included a power monitor, enabling VOA~2 to be adjusted to match the optical fibre loss.
After transmission over fibre or VOA~2, a pre-amplifier and the VOA~3 were utilised to control the signal power into the receiver.  
The signal was detected by a 70~GHz coherent receiver frontend to convert the optical signal into the electrical domain and captured on a 10-bit, 256 GSa/s real-time Keysight UXR oscilloscope, followed by pilot-based digital signal processing (DSP) described in~\cite{wakayama20212048}.

The optical fibres investigated included the TXF and Vascade EX2500, with attenuation values at 1490~nm of 0.185~dB/km and 0.168~dB/km, respectively, and an effective area of 125~\textmu m$^2$.
The fibres were selected to have CC at wavelengths at the upper end of the manufacturing distribution.
The average CC for the TXF fibre is slightly less than 1500~nm and that of EX2500 fibre is approximately 1480~nm.
The primary metric used to estimate transmission performance was the average SNR measured from five received traces.
Additionally, the OSNR metric was considered, with a 0.1~nm reference bandwidth, to compare received signals, as it is only influenced by signal power and ASE power over the signal bandwidth.
The signals were transmitted over distances of approximately 80~km and 160~km for each fibre type, with small variations in lengths due to the availability of spools.
The launch power, for each case, was selected to maximise the received SNR in the linear regime.
This approach ensured negligible nonlinear interference noise, such that the most impactful sources of impairments in the measured SNR were the transceiver (TRX) noise, ASE noise from TDFAs, and the potential MPI, given that the DSP fully compensated chromatic dispersion.
Table from Fig.~\ref{fig:setup} shows the TXF and Vascade EX2500 fibre transmission configurations that were analysed and compared with a signal obtained using VOAs and TDFAs only.
The transmission was carried out for 22 channels ranging from 1474~nm to 1525~nm, for each combination of baud rate, transmission distance, and fibre type.

The transmission performance of TXF and Vascade EX2500 fibres were also compared with that of a 161.3~km G.652.D fibre span with 80~\textmu m$^2$ effective area and attenuation of 0.21~dB/km at 1490~nm using different baud rates (see inset of Fig.~\ref{fig:setup}).
Note that the experimental bend diameter condition on shipping reels is similar to those in a loose-tube cable, representing a realistic scenario for field-installed~fibre.
 
\begin{figure*}[t]
    \centering
    \includegraphics[width = 0.92\textwidth]{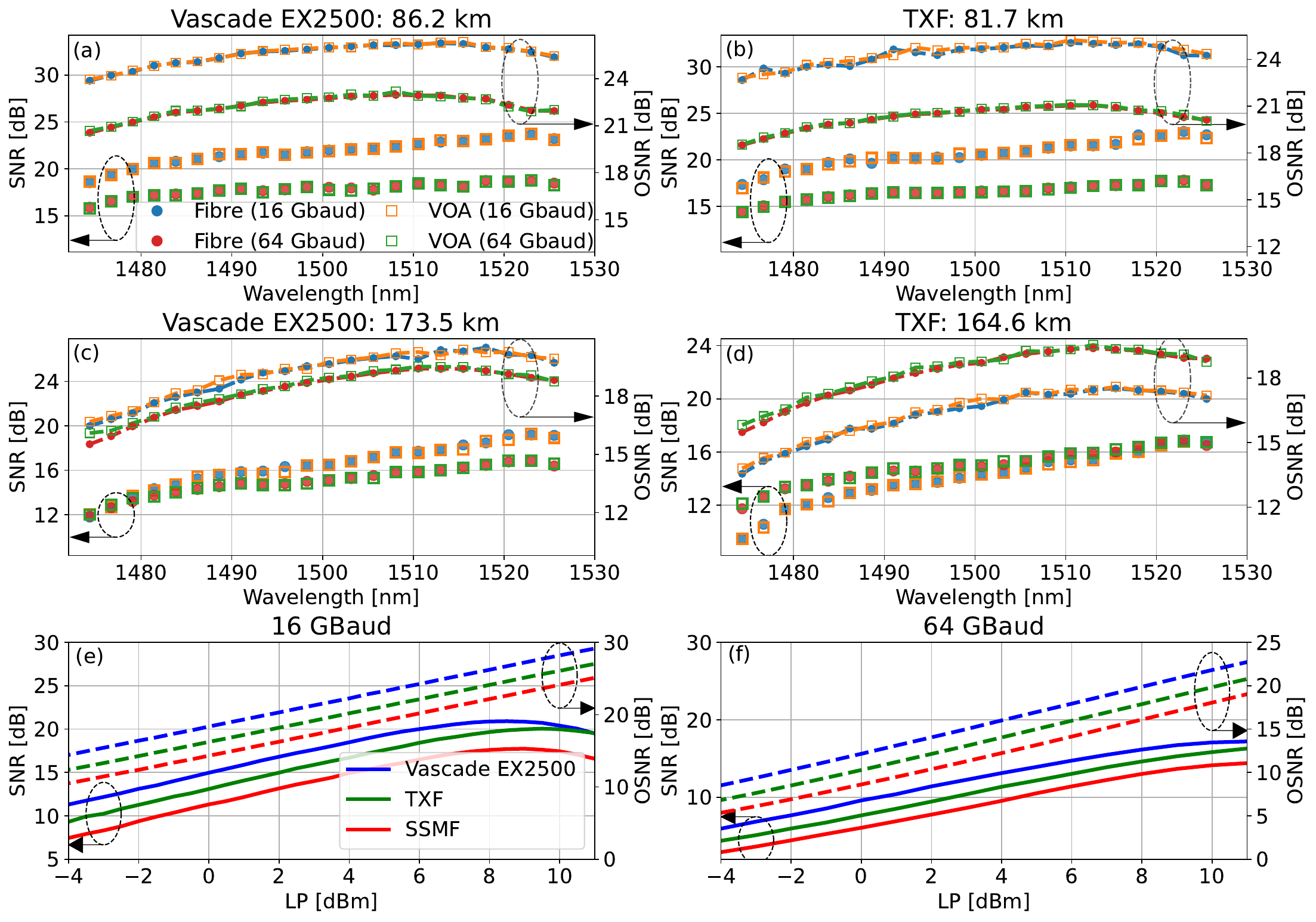}
    \setlength{\belowcaptionskip}{-0pt}
    \caption{Measured SNR and OSNR for transmission over VOA and fibre with baud rates of 16~GBaud and 64~GBaud for (a) 80~km Vascade EX2500, (b) 80~km TXF, (c) 160~km Vascade EX2500, and (d) 160~km TXF. Filled circle and hollow square markers indicate fibre transmission over fibre and VOA, respectively; the OSNR is shown as dashed lines. Measured SNR and OSNR against LP over approximately 160~km of Vascade EX2500, TXF, and G.652 at 1490~nm for (e) 16~GBaud, and (f) 64~GBaud. Continuous lines indicate SNR and dashed lines indicate OSNR.}
    \label{fig:results}
\end{figure*}

\vspace{-2pt}

\section{Experimental Results and Discussion}

\vspace{-0.15cm}

The SNR and OSNR were measured for the 22 channels in the S-band after transmission over Vascade EX2500 and TXF fibres, for the lengths and baud rates listed in Fig.~\ref{fig:setup}.
The transmission results are compared with the performance over the VOA~2 in place of the fibre span and are shown in Fig.~\ref{fig:results}(a)-Fig.~\ref{fig:results}(d).
It can be seen that the SNR values measured after transmission over fibre predominantly overlap with those for the VOA~2 case for every baud rate and transmission distance, which indicates that the ASE and TRX noise are the only significant noise sources causing SNR degradation, while the MPI effect is negligible and does not further impair the SNR in all transmission scenarios.
SNR mismatches between VOA and fibre transmission lower than 0.4~dB were measured for every channel in all transmission configurations.
The maximum SNR discrepancy between the fibre and VOA transmission was observed for the 64~GBaud 160~km TXF fibre case at 1474~nm, where the SNR was 0.39~dB lower than the VOA~2 case.
However, this SNR mismatch corresponds to a similar OSNR difference of 0.33 dB between the VOA and fibre transmissions, suggesting that it may result from LP variations due to the polarisation state of the light entering the modulator or from a variation in the modulator bias.

To further assess the advantages of low-attenuation and high effective area of G.654.E fibres over the G.652.D fibre with 80~\textmu m$^2$ effective area, we measured received SNR and OSNR for the three fibres mentioned in the Fig~\ref{fig:setup} inset table.
Figure~\ref{fig:results}(e) shows the SNR measurements at 1490~nm for 16~GBaud, and Fig.~\ref{fig:results}(f) shows the same results but for a symbol rate of 64~GBaud. 
Vascade EX2500, TXF, and G.652.D spool lengths were 173.5~km, 164.6~km, and 161.3~km, respectively, resulting in attenuation of 29.15~dB, 30.45~dB, and 33.87~dB. 
Notably, Vascade EX2500 demonstrated approximately 2~dB higher OSNR than the TXF, while TXF showed a similar gain over G.652.D fibre. 
This gain in the OSNR (resulting from the lower attenuation) is reflected in the SNR for LP values in the linear regime. 
For the case of high values of LP, the SNR starts to reduce due to nonlinearities.
The very low loss of Vascade EX2500 fibre increases its effective length, thus resulting in a lower optimum LP compared to the G.652.D fibre, however presenting a significantly higher optimum SNR.
The maximum SNR values measured for Vascade EX2500, TXF, and G.652.D fibres were 20.91~dB, 20.05~dB, and 17.74~dB, respectively, corresponding to optimum LP levels of 8.5~dBm, 9.5~dBm, and 9~dBm for a baud rate of 16~GBaud.
For the 64~GBaud scenario, TDFA gain limitations capped LP at 11~dBm and determining optimal LP for 64~GBaud was not feasible - for this case, SNR values were 17.2~dB, 16.3~dB, and 14.4~dB, at LP of 11~dBm.

\vspace*{-0.2cm}

\section{Conclusions}

\vspace{-0.15cm}


We demonstrated MPI-free transmission using S-band signals over G.654.E-compliant fibres, which feature cable cutoff wavelengths higher than the S-band range. 
SNR comparisons were performed for 22 channels (1474-1525~nm) over VOA, Vascade EX2500 (average CC: 1500~nm), and TXF fibres (average CC: 1480~nm) at distances of 80~km and 160~km, and for 16~GBaud and 64~GBaud. 
Comparable SNR values were measured for G.654.E fibres and VOA, with variations lower than 0.4~dB, attributed to similar OSNR fluctuations.
Additionally, an SNR improvement of more than 3 dB was achieved when S-band channels were transmitted over Vascade EX2500 compared to G.652.D fibre, over similar distances and with lower optical launch power.
Thus, G.654.E-compliant fibres demonstrate suitability for S-band transmission with negligible MPI penalties and significant performance enhancement from their lower attenuation.

\vspace{2pt}
\noindent\textbf{Acknowledgements:} This work was supported by EPSRC grants EP/R035342/1 TRANSNET
, EP/W015714/1 EWOC
, and EP/V007734/1 EPSRC Strategic Equipment Grant.

\vspace{-2pt}

\begin{thebibliography}{99}
\footnotesize
\vspace{-12pt}

\begin{multicols}{2}

\bibitem{Winzer:18} P. J. Winzer, et al., Opt. Express \textbf{26}(18), pp. 24190-24239 (2018)

\bibitem{bayvel2016maximizing} P. Bayvel, et al., Phil. Trans. R. Soc. A \textbf{374}, 20140440 (2016).







\bibitem{9893169} T. Hoshida, et al., Proc. IEEE \textbf{110}(11), pp. 1725-1741 (2022).

\bibitem{Puttnam:22} B. J. Puttnam, et al., in \emph{OFC}, paper W3C.5 (2022).

\bibitem{9489895} B. J. Puttnam, et al., in \emph{OFC}, paper Th4C.2 (2021).

\bibitem{9144561} L. Galdino, et al., IEEE \emph{PTL} \textbf{32}(17), pp. 1021-1024 (2020).

\bibitem{Maeda:21} H. Maeda, et al., \emph{JLT} \textbf{39}(4), pp. 933-939 (2021).

\bibitem{Mlejnek:19} M. Mlejnek, et al., \emph{JLT} \textbf{37}(17), pp. 4282-4294 (2019).


\bibitem{9748243} V. V. Ivanov, et al., in \emph{OFC}, paper W3E.4 (2022).

\bibitem{8345898} J. D. Downie, et al., in \emph{ECOC}, paper P1.SC1.6 (2017).

\bibitem{101109ecoc2021} H. Li, et al., in \emph{ECOC}, paper We3C1.5 (2021).

\bibitem{itut} ITU-T, Rec. G.654 (2020). [Online].


\bibitem{downie2023ecoc} J. D. Downie, et al., in \emph{ECOC}, paper We.C.2.5 (2023).

\bibitem{Lin:21} J. Lin, et al., in \emph{ACP}, paper T4A.107 (2021).

\bibitem{wakayama20212048} Y. Wakayama, et al., Opt. Express \textbf{29}(12), pp. 18743-18759 (2021).







\end{multicols}
\end{thebibliography}

\end{document}